\documentclass[preprint,NumberedRefs]{JASAnew}
\usepackage{siunitx} 

\begin{document}

\title[JASA/Actuation of needles for drug delivery]{An Ultrasonically Actuated Needle Promotes the Transport of Nanoparticles and Fluids}

\author{Emanuele Perra}
\author{Nick Hayward}
\affiliation{Medical Ultrasonics Laboratory (MEDUSA), Department of Neuroscience and Biomedical Engineering, Aalto University, Espoo, 02150, Finland}{}

\author{Kenneth P.H. Pritzker}
\affiliation{Department of Laboratory Medicine and Pathobiology, University of Toronto, Toronto, M5S 1A8, Canada}
\affiliation{Department of Pathology and Laboratory Medicine, Mount Sinai Hospital, Toronto, M5G 1X5, Canada}

\author{Heikki J. Nieminen}
\email{heikki.j.nieminen@aalto.fi}
\thanks{Corresponding Author}
\affiliation{Medical Ultrasonics Laboratory (MEDUSA), Department of Neuroscience and Biomedical Engineering, Aalto University, Espoo, 02150, Finland}

\preprint{Emanuele Perra, JASA}		%  if you want want this message to appear in upper left corner of title page

\date{\today} 

\begin{abstract}
Non-invasive therapeutic ultrasound methods, such as high-intensity focused ultrasound (HIFU), have limited access to tissue targets shadowed by bones or presence of gas. This study demonstrates that an ultrasonically actuated medical needle can be used to translate nanoparticles and fluids under the action of nonlinear phenomena, potentially overcoming some limitations of HIFU. A simulation study was first conducted to study the delivery of a tracer with an ultrasonically actuated needle (\SI{33}{\kHz}) inside a porous medium acting as a model for soft tissue. The model was then validated experimentally in different concentrations of agarose gel showing a close match with the experimental results, when diluted soot nanoparticles (diameter $<$ \SI{150}{\nm}) were employed as delivered entity. An additional simulation study demonstrated a threefold increase of the volume covered by the delivered agent in liver under a constant injection rate, when compared to without ultrasound. This method, if developed to its full potential, could serve as a cost effective way to improve safety and efficacy of drug therapies by maximizing the concentration of delivered entities within \textit{e.g.} a small lesion, while minimizing exposure outside the lesion.
\end{abstract}

\maketitle

\section{Introduction}
During recent decades, high-intensity ultrasound (HIU) methods for actuation have been broadly investigated. In fact, the capability of delivering a considerable amount of acoustic energy into small targets and the ability to actuate matter from distance is of interest to a number of medical applications. HIU waves carry non-ionizing radiation, therefore mitigating safety concerns and, thus, allowing variety of treatment strategies \cite{Ebbini2015}.

High-intensity focused ultrasound (HIFU), an application of HIU, permits deposition of thermal energy to a focal volume leading to \textit{e.g.} ablation \cite{Zhou2011}, hyperthermia \cite{Draper1995} or drug release from thermally sensitive drug vehicles such as thermosensitive liposomes (TSL) \cite{Gasselhuber2012}. Primarily non-thermal applications of HIFU include sonoporation \cite{Mehier-Humbert2005}, sonophoresis \cite{Vranic2004} and drug transportation \cite{Pitt2004}. However, limitations of this approach include unwanted biological effects potentially induced on the tissue, \textit{e.g.} thermal effects caused by the ultrasound absorption of biological tissues, thermal and mechanical effects induced by cavitation \cite{OBrien2007}, and the difficulty to generate a proper focus at certain anatomical regions due to the acoustic shadowing of bones and presence of gas in the respiratory system \cite{Kim2008}. For example, tumors located at the hepatic dome are difficult to treat, because the right lower lobe of the lung, containing air preventing sound propagation, partially covers this area, hence limiting targeted delivery of acoustic energy. In order to overcome this issue, the artificial pleural effusion medical procedure has been widely used to facilitate the HIFU treatment of tumors in the hepatic dome. However, this procedure is highly invasive, risky, and can lead to complications, thus it requires an experienced operator \cite{Long2020}. Also, the reflected acoustic energy may cause skin burns if uncontrolled \cite{Zhu2009}, for which reason the invasive procedure of partial rib resection has even been recently proposed and evaluated as a way to create an acoustic pathway for HIFU treatments \cite{Long2020}. Moreover, the translation of organs during breathing cycles can impact on the precision, safety and efficiency of the HIFU treatment methods \cite{Tharkar2019a}.

Given these limitations, percutaneous injections \textit{via} hypodermic needles still provide an alternative and cost-effective technique to administer drugs, large molecules, nano- and microparticles for the treatment of pathologies within organs that are hardly accessible by ultrasound. However, considering the importance and extensive use of hypodermic needles, limited research has been conducted to explore how the combination of medical needles and nonlinear ultrasonics (NLU) could add value to medical applications such as enhanced drug delivery. Conventional approaches such as localized injection techniques rely on spreading the therapeutic agent uniformly and in a great volume within the target. However, most of the percutaneous injection techniques have limitations such as the low absorption rate of the substance and the restricted amount of drug that has to be administered, often resulting in limiting the therapeutic effect \cite{Amalou2013}; thus the improvement of the current needle designs and functionality must be considered \cite{Sudheendra2006}. New ways for delivery may be beneficial for treatment of soft tissue pathologies with localized and targeted strategies \cite{precisionmedicine} in liver as well as other anatomical locations.

In this study, we investigate the capability of an ultrasonically actuated medical needle \cite{Perra2021} to translate nanoparticles and fluids in soft tissue under the action of NLU phenomena. The hypodermic needle, when used in combination with a pressure source such as a syringe, is employed to act as a fluid conduit. This allows operators to control the deposit of high volumes of fluid into the target location. When actuated with ultrasound (US), the needle also serves as a waveguide, which conducts the ultrasonic power to the target \cite{Perra2021}. We, therefore, aim to demonstrate how acoustic radiation force (ARF) can act on the liquid to enhance convective transport, thus allowing influencing drug distribution within a target (Fig. \ref{figure1}a). Numerical soft tissue modeling is used to simulate the acoustic and flow fields generated by the ultrasonic needle in a porous medium. The numerical results are then validated experimentally by comparing the penetration front distribution of the tracer over time in different concentrations of agarose gels with similar settings as those from the simulation environment. The potential of using US in combination with a hypodermic needle for the delivery of fluids and entities is finally studied numerically in a liver tissue model, exploring its potential in improving percutaneous ethanol injection (PEI) for the treatment of liver cancer.  

\section{Methods}
\subsection{Numerical Simulations}
\begin{figure*}[ht]
\includegraphics[width=\textwidth]{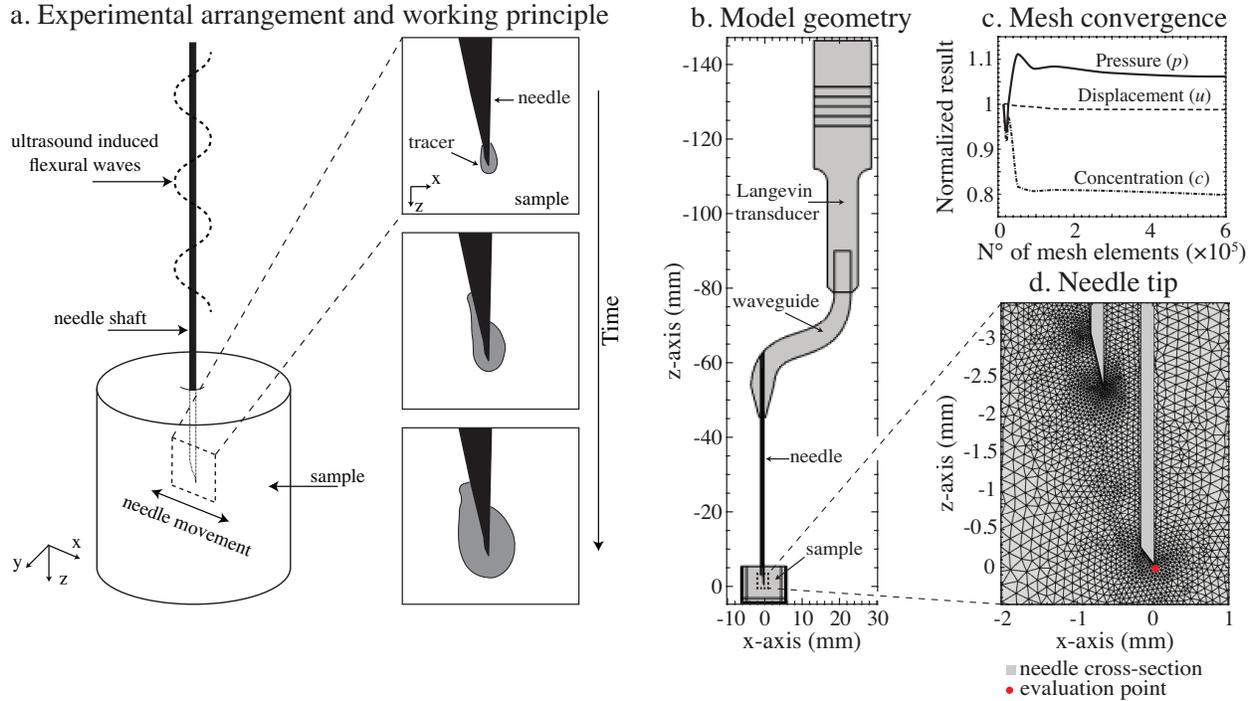}
\caption{\label{figure1}{(\textbf{a}) Schematics representing the working principle of the concept. The hypodermic needle is actuated with ultrasound to induce flexural waves while the needle tip is inserted into the sample. The ultrasonic action is meant to facilitate the delivery of the substance injected over time. (\textbf{b}) Cross-section of the three-dimensional model used in the numerical simulations. The model replicates the actual ultrasonic device used in the experiments, comprising of a Langevin transducer that is coupled to a hypodermic needle through an S-shaped aluminum waveguide. (\textbf{c}) A mesh convergence study has been conducted to determine an adequate mesh resolution around the needle tip region. Multiple simulations were carried out by decreasing the element size and the convergence of different physical variables was evaluated on (\textbf{d}) a mesh node next to the needle tip (highlighted in red). Based on the convergence study, an element size in the range 0.027-\SI{0.416}{mm} was selected from the sample domain.}}
\end{figure*}

Fig. \ref{figure1}b-d represents the physical model adopted in the simulations. The schematics depicted in Fig. \ref{figure1}b shows a cross section of the 3D replicate of the device used in the experiments, comprising of a Langevin transducer, an S-shaped waveguide and a 21G $\times$ \SI{80}{mm} hypodermic needle. The needle tip is partially placed into a \SI{10}{mm} $\times$ \SI{12}{mm} cylinder which is considered to be a porous medium, where the different equations are solved by using the computational software COMSOL Multiphysics v5.5 \cite{Multiphysics2020}.

The different physics involved in the fine element model (FEM) are the following: electrostatics in the piezoelectric stack and its respective stress-charge constitutive relation; elastodynamics in all domains except for the sample domain; acoustics, fluid dynamics and solute transport within the sample domain. First, the equation of motion within the ultrasonic device and the acoustic pressure field inside the sample domain are solved in the frequency domain by applying a potential difference of \SI{15}{V} across the faces of the piezoelectric rings. The structural acceleration of the portion of the needle tip is used as a boundary condition for the calculation of the acoustic pressure across the needle-sample interface as in the following mathematical condition: $-\textbf{n}\cdot \frac{1}{\rho}\nabla p = \textbf{n}\cdot \textbf{u}_{tt}$, where $\textbf{u}_{tt}$ is the structural acceleration of the moving interface, $p$ is the acoustic pressure, $\rho$ is the density of the medium and $\textbf{n}$ is the vector normal to the boundary interface. This condition state that, given a positive acceleration of the interface, a negative pressure gradient is generated along the positive direction defined by $\textbf{n}$. The stationary fluid flow within the porous domain is then evaluated by solving the Brinkman equations \cite{Brinkman1949}. In order to couple the acoustics domain to the equations of fluid motion, the Reynolds stress, which arises from the sound attenuation in the fluid and is responsible for generating streaming \cite{Zarembo1971}, is considered as the force term in such equations. Finally, the equations governing the transport of diluted species inside a porous medium are solved in the time domain. In this step, in order to account for the solute transport by convection, the solution of the Brinkman equations is considered to be the background flow velocity field. 

Free tetrahedral elements were used to mesh the three-dimensional model, considering at least 20 nodes per wavelength according to the speed of wave propagation in each material. The number of nodes per wavelength was considered appropriate to minimize the local approximation errors \cite{Thompson2006}. Since we were interested in analysing the acoustic-related phenomena at the needle tip, the mesh resolution was increased in this region, and a mesh convergence study was performed to determine its optimal mesh size. It was found that an element size in the range 0.027-\SI{0.416}{mm} in the sample domain was appropriate to obtain accurate and consistent results (Fig. \ref{figure1}c). A detailed list of the model parameters is given in Table \ref{Table1}. 

In the following sections, the mathematical equations used in the numerical simulations are described.

\subsubsection{Acoustic Wave Propagation}
The tissue is assumed to be a viscous fluid. The acoustic propagation in a viscous fluid is described by the viscoelastic wave equation \cite{Holm2011}:
\begin{equation}
\begin{aligned}
&\nabla^2 p -\frac{1}{c_0^2}\frac{\partial^2p}{\partial t^2} + \frac{\delta}{c_02}\frac{\partial}{\partial t}\nabla^2 p= 0,
\end{aligned}
\label{viscous}
\end{equation}
where $c_0$ is the speed of sound (\SI{}{\m\per\s}) and $\delta$ is the sound diffusivity (\SI{}{\square\meter\per\s}). The first two terms of Eq. (\ref{viscous}) describe the linear lossless propagation of sound in a medium, while the third term is associated with viscous losses. Assuming an isotropic material, a general solution for Eq. (\ref{viscous}) is:
\begin{equation}
\begin{aligned}
& p(\vec{x},t) = p_0e^{j(\omega t- k_x\cdot x-k_y\cdot y-k_z\cdot z)}
\end{aligned}
\label{gensol}
\end{equation}
which, if substituted into Eq. (\ref{viscous}), leads to the follow equation
\begin{equation}
\begin{aligned}
& k_x^2 + k_y^2 + k_z^2 = |k|^2 = \frac{\omega^2}{c_0^2 (1+\frac{j\omega\delta}{c_0^2})}.
\end{aligned}
\label{kvec}
\end{equation}
The time-harmonic representation of Eq. (\ref{viscous}) is:
\begin{equation}
\begin{aligned}
&\nabla^2 p+ k_{eq}^2p = 0,\\
& k_{eq}^2 = \Big(\frac{\omega}{c_c}\Big)^2,\\
& c_c = c_0\Big(1+ \frac{i\omega\delta}{c_0}\Big)^{0.5},
\end{aligned}
\label{acwave}
\end{equation}
where $\omega$ is the angular frequency and $c_0$ is the speed of sound. The sound diffusivity $\delta$, which accounts for viscous losses in a viscous fluid, is modeled as \cite{Solovchuk2013,Demi2014}:
\begin{equation}
\begin{aligned}
&\delta = \frac{2c_0^3\alpha}{\omega^2},
\end{aligned}
\label{diffusivity}
\end{equation}
where $\alpha$ denotes the sound attenuation coefficient in a specific medium.

\subsubsection{Fluid Flow in Porous Medium}
The fluid movement inside an agarose gel or a soft tissue is modeled as an incompressible and steady flow in a porous medium fully saturated with water. The equation of motion and continuity of the fluid based on the Brinkman model are \cite{Brinkman1949}:
\begin{equation}
\begin{aligned}
& \nabla p - \frac{\mu}{\epsilon_p}\nabla^2\textbf{u}+\frac{\mu}{k}\textbf{u} = \textbf{F},\\
&\nabla \cdot \textbf{u} = 0,
\end{aligned}
\label{brinkmann}
\end{equation}
where $p$ denotes the pressure, $\mu$ viscosity, $\epsilon_p$ porosity, $\textbf{u}$ the velocity vector, $k$ permeability and $\textbf{F}$ the force term. The acoustic streaming \cite{Nyborg1965} is assumed to be the main cause of fluid motion in the tissue interstitial spaces. Therefore, the force term $\textbf{F}$ accounting for the momentum transfer from the acoustic wave to the fluid is given by the following time-averaged volume force:
\begin{equation}
\begin{aligned}
\textbf{F} = \langle \rho \frac{\partial \textbf{v}}{\partial t} \rangle + \rho \langle (\textbf{v} \cdot \nabla)\textbf{v}\rangle ,
\end{aligned}
\label{acstream}
\end{equation}
where $\rho$ is the density of the fluid and $\textbf{v}$ is the acoustic velocity. As the time average of a quadratic periodic variable is non-zero, the magnitude of the driving force $\textbf{F}$ responsible for the acoustic streaming generation will be always nonzero, when an acoustic wave is travelling in a fluid. 

\subsubsection{Permeability and Porosity}
The porosity of the agarose gel was calculated as \cite{Johnson1996,Pluen1999}:
\begin{equation}
\begin{aligned}
& \epsilon_p = 1-\phi,
\end{aligned}
\label{porperm}
\end{equation}
where $\phi$ is the volume fraction of the agarose fiber expressed by: 
\begin{equation}
\begin{aligned}
\phi = c_{agar}/\rho_{agar} \omega_{agar}
\end{aligned}
\label{volumfrac}
\end{equation}
In Eq. (\ref{volumfrac}) $c_{agar}$ is the agarose concentration (w/v) in the gel, $\rho_{agar}$ is the density of dry agarose (\SI{1.64}{\gram\per \mL} \cite{Laurent1967}) and $\omega_{agar}$ is the agarose mass fraction in a fiber (0.625), determined experimentally by Johnson \textit{et al.} \cite{Johnson1995a}.
The Carman-Kozeny equation was used to determine the hydrogel permeability, $\kappa$, as follows \cite{Debbaut2014}: 
\begin{equation}
\begin{aligned}
\kappa = \frac{\epsilon_p r_h^2}{k}
\end{aligned}
\label{perm}
\end{equation}
The hydraulic radius was assumed to be similar to the inter-fiber spacing, $r_h\approx \bar a/2$, and the average gel pore size, $\bar a$, was determined experimentally by Narayanan \textit{et al.} \cite{Narayanan2006a} as a function of the agarose concentration.
The Kozeny factor, $k$, takes into account the direction of the gel interstitia considered as cylindrical pores randomly oriented in the 3D space. For hydrogels with high void volumes ($\epsilon > 0.9$) it is defined as \cite{Levick1987}:
\begin{equation}
\begin{aligned}
& k = (2k_+ + k_{|})/3,\\
& k_+ = \frac{2\epsilon^3}{{(1 - \epsilon) \cdot [\text{ln}(\frac{1}{1-\epsilon})-\frac{1-(1-\epsilon)^2}{1+(1-\epsilon)^2}]}},\\
& k_{|} = \frac{2\epsilon^3}{{(1 - \epsilon)\cdot[2\text{ln}(\frac{1}{1-\epsilon})-3+4\cdot(1-\epsilon)-(1-\epsilon)^2]}},
\end{aligned}
\label{kozeny_factor}
\end{equation}
where $k_+$ and $k_{|}$ account for cylindrical interconnected pores perpendicular and parallel to the fluid flow, respectively. 
Tissue permeability in the liver can be estimated by considering the tissue as comprised of spherical objects organized into a grid-like structure. Each sphere has a similar diameter to one hepatocyte (\SI{24}{\micro m}). By using the Carman-Kozeny model for flow around a spherical object \cite{Dullien1979}, the permeability is calculated as follows:
\begin{equation}
\begin{aligned}
\kappa = \frac{d^2(1-\epsilon)^3}{180\epsilon^2},
\end{aligned}
\label{permkozenyliver}
\end{equation}
where $d$ is the diameter of the spherical object, and $\epsilon$ is the porosity of the tissue. The porosity can be easily obtained by calculating the void ratio in a cell containing a sphere with radius $R$ encapsulated in a cube with length $2R$, which represents the repeating cell unit in the grid-like structure. This gives:
\begin{equation}
\begin{aligned}
\epsilon = 1- \frac{V_{sphere}}{V_{cube}} = 1- \frac{4\pi (0.5)^3}{3} = 0.476.
\end{aligned}
\label{porosityliver}
\end{equation}

\subsubsection*{Solute Transport in Porous Medium}
The equation governing the diffusion of solute species in an isotropic porous medium is given by \cite{Quastel1998}:
\begin{equation}
\begin{aligned}
\epsilon_p\frac{\partial c}{\partial t} +\textbf{u}\cdot\nabla c- D_{eff} \nabla^2 c=0,
\end{aligned}
\label{diffusion}
\end{equation}
where $c$ is the concentration of the tracer ($\SI{}{\mole\per\m^3}$), $\textbf{u}$ the velocity field derived from Eq. (\ref{brinkmann}) and $D_{eff}$ is the effective diffusion coefficient of the tracer in a porous medium. This is defined as:
\begin{equation}
\begin{aligned}
 D_{eff} = \frac{D_0\epsilon_p}{\tau},
\end{aligned}
\label{effectivediff}
\end{equation}
where $D_0$ is the diffusion coefficient of ink in water (\SI{2.5e-10}{\square\meter\per\second} at \SI{20}{\celsius} \cite{Lee2004}), $\epsilon_p$ is the porosity and $\tau$ is the Millington-Quirk tortuosity coefficient \cite{Ray2019}, which is expressed as a function of the porosity as follows: $\tau = \epsilon_p^{-1/3}$.

\subsection{Experiments}

\subsubsection{Hydrogel Preparation}
Hydrogels with agarose concentrations of 0.5, 1 and 2\% (w/v agarose powder/EDTA) were prepared by adding respectively 0.25, 0.5 and \SI{1}{\g} of agarose (catalogue number: 10377033, Agarose Low-Melting, Nucleic Acid Recovery/Molecular Biology Grade, Thermo Fisher Scientific, Waltham, MA, United States) to \SI{50}{\milli\liter} of \SI{50}{\milli M} EDTA buffer (catalogue number: 11836714, Thermo Scientific EDTA, Thermo Scientific, Waltham, MA, United States). The solution was heated up to \SI{96}{\degreeCelsius}, while being mixed with a magnetic stirrer (Heater and magnetic stirrer C-MAG HS series C-MAG HS 4 model, IKA, Staufen, Germany), and allowed to cool down at room temperature (\SI{22}{\degreeCelsius}). When the temperature of the solution had reached \SI{42}{\degreeCelsius}, \SI{3}{\milli\liter} of agarose gel was poured into different polystyrene cuvettes (external dimensions = L $\times$ W $\times$ H = \SI{12}{mm} $\times$ \SI{12}{mm} $\times$ \SI{45}{mm}, wall thickness = \SI{1}{mm}) and left to solidify for 2 hours.

\subsubsection{Sonication in Agarose}
A custom-made ultrasonic device \cite{Perra2021} was employed to enable flexural waves in a hypodermic needle and mediate the transport of the tracer in different agarose concentrations. The design consisted of a Langevin transducer coupled to a hypodermic needle (21G, length = $\SI{80}{\mm}$) (model: 4665465, 100 STERICAN, B Braun, Melsungen, Germany) \textit{via} an S-shaped 3D printed aluminum waveguide (3D Step Oy, Ylöjärvi, Finland). The needle tip was first moistened with 40\% (v/v ink/deionized water) diluted ink (Winsor \& Newton Calligraphy Ink Bottle, 30ml, Black, Winsor \& Newton, London, United Kingdom) and then inserted into the agarose specimen at a depth of \SI{3}{mm} with an insertion velocity of \SI{50}{\micro\meter\per\second}, which was controlled by a motorized three-axis translation stage (model: 8MT50-100BS1-XYZ, Motorized Translation Stage, Standa, Vilnius, Lithuania).
Continuous waves at the frequency of \SI{33}{\kHz} were applied to the needle for a duration of \SI{10}{\minute} by using an RF amplifier (model: AG 1012LF, Amplifier/Generator, T\&C Power Conversion, Inc., Rochester, NY, United States) in combination with a function generator (model: Analog Discovery 2, Digilent, Inc., Henley Court Pullman, WA, United States). The needle action was imaged throughout the duration of the experiment with a high-speed camera (model: Phantom V1612, Vision Research, Wayne, NJ, United States) in conjunction with a macro lens (model: Canon MP-E 65 mm f / 2.8 1-5x Macro Photo, Canon Inc., Ōta, Tokyo, Japan) using the following settings: sample rate = \SI{100}{fps}, exposure = \SI{9900}{\micro\second}, resolution = 768 pixels × 768 pixels, lens aperture = 16. Eventually, the projected area of the tracer distribution over time was quantified and analyzed in MATLAB (R2020b) \cite{Mathworks2016} from the recorded images as follows:
\begin{equation}
\begin{aligned}
 A_{tracer, i} = \int\int I_{bw,i}(x,y)- I_{bw,1}(x,y) \,dx\,dy,
\end{aligned}
\label{effectivediff}
\end{equation}
where $I_{bw,i}(x,y)$ are binary frames generated by thresholding the frame-set $I(x,y)$ with the Otsu method \cite{Otsu1996} and $I_{bw,1}(x,y)$ is the first frame of the video footage in which all needle boundaries are clearly visible.

\section{Results}
Fig. \ref{figure1}b represents the geometry of the three dimensional model employed in the simulations. In this design, the waveguide converts the longitudinal motion provided by the transducer to a flexural movement of the needle. This creates flexural standing waves in the needle with greatest displacement at its tip. By this means, the needle tip is made to oscillate sideways within the xz-plane, acting as a dipole-like sound source. The emitted sound is anticipated to exert an acoustic radiation force on the liquid, promoting the convection of fluid through the interstitia of the porous medium. The model was validated by comparing the tracer delivery dynamics calculated numerically, \textit{i.e.} cross-sectional area of tracer distribution, expansion velocity of the penetration front and tracer concentration profiles, to those measured experimentally, \textit{i.e.} projected area of tracer distribution, expansion velocity of the penetration front and tracer absorbance profiles. 

\subsection{Characterization of Delivery in Different Agarose Gel Concentrations}
\begin{figure*}[ht]
\includegraphics[width=\textwidth]{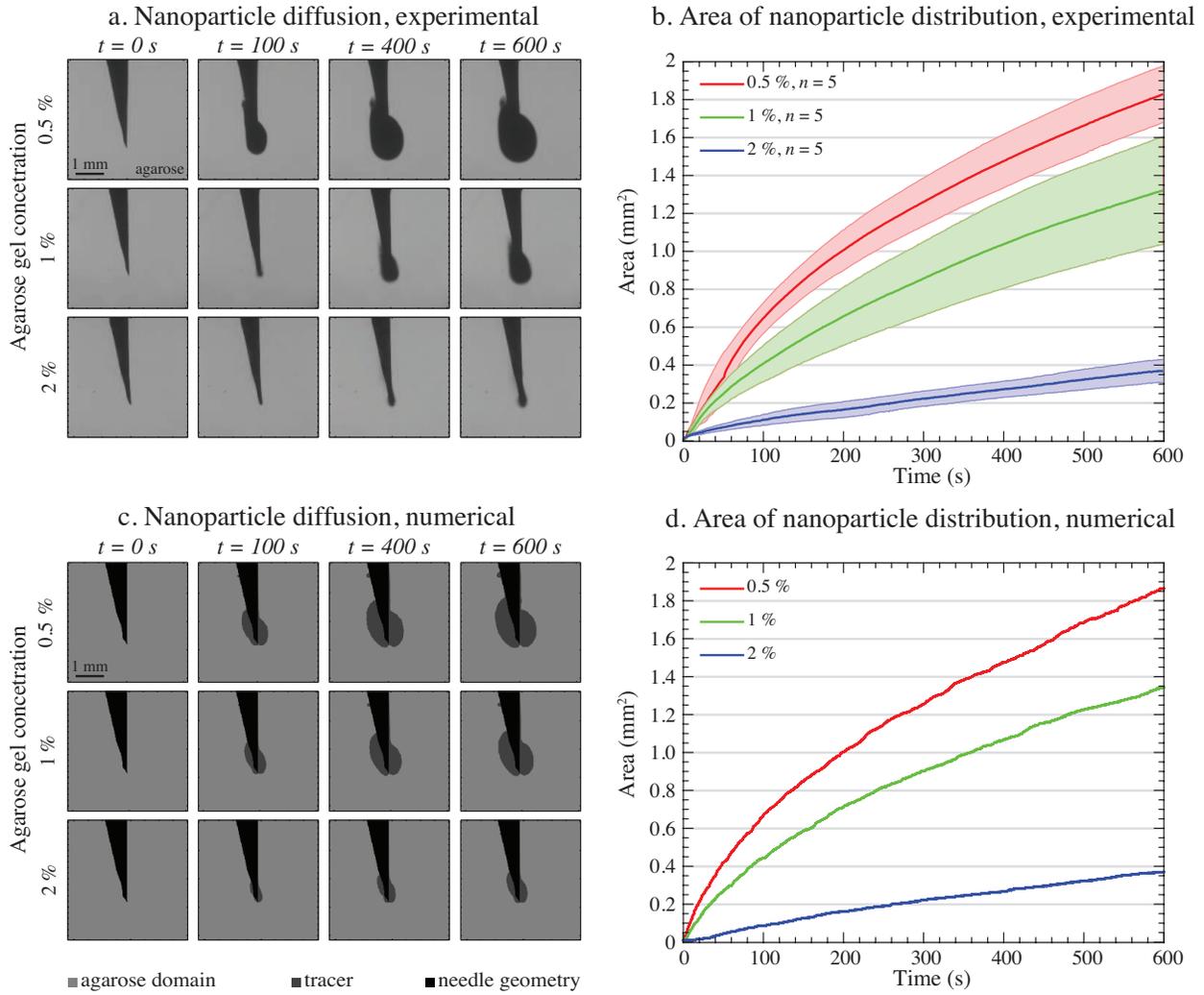}
\caption{\label{figure2}{(\textbf{a},\textbf{b}) Shadowgraphy of the delivery effect at different time points and in different concentrations of agarose. When 40\% (v/v, ink/deionized water) diluted ink was used as a tracer, the area of delivery increased with time and decreased at lower concentrations of agarose. In \textbf{b}, the solid lines represent the mean value of the data within the same experimental group and the shaded areas indicate the 95\% confidence interval. (\textbf{c},\textbf{d}) The experiments were in line with the simulation results.}}
\end{figure*}

Hydrogels with agarose concentration of 0.5, 1 and 2\% (w/v agarose/EDTA) were subjected to \SI{10}{min} of sonication applied with the ultrasonic needle, while the needle tip was moistened with 40\% (v/v ink/deionized water) and inserted \SI{5}{mm} deep into the specimen. The projected area of the spatial distribution of the tracer was imaged throughout the experiment and quantified during the post processing. Fig. \ref{figure2}a represents different time frames of the ultrasonic mediated delivery of nanoparticles in hydrogels acquired at 0, 100, 400 and \SI{600}{s} from the beginning of the sonication. The projected area of nanoparticle distribution increased over time, covering an area of \SI{1.85}{mm^2} on average after \SI{10}{min} in 0.5\% agarose gel (Fig. \ref{figure2}b). As the gel concentration increases, the area of distribution decreases, measured to be 1.35 and \SI{0.35}{mm^2} in 1 and 2\% agarose gel, respectively. These results were in line with the numerical simulations, where the cross section of the three dimensional model exhibited pronounced distribution of nanoparticles around the needle tip, which grew over time (Fig. \ref{figure2}c). Similarly to the experimental observations, the delivery of nanoparticles (diameter $<$ \SI{150}{nm}) was confined to a smaller region, when higher concentrations of agarose gel were used (Fig. \ref{figure2}d).

\begin{figure*}[ht]
\includegraphics[width=\textwidth]{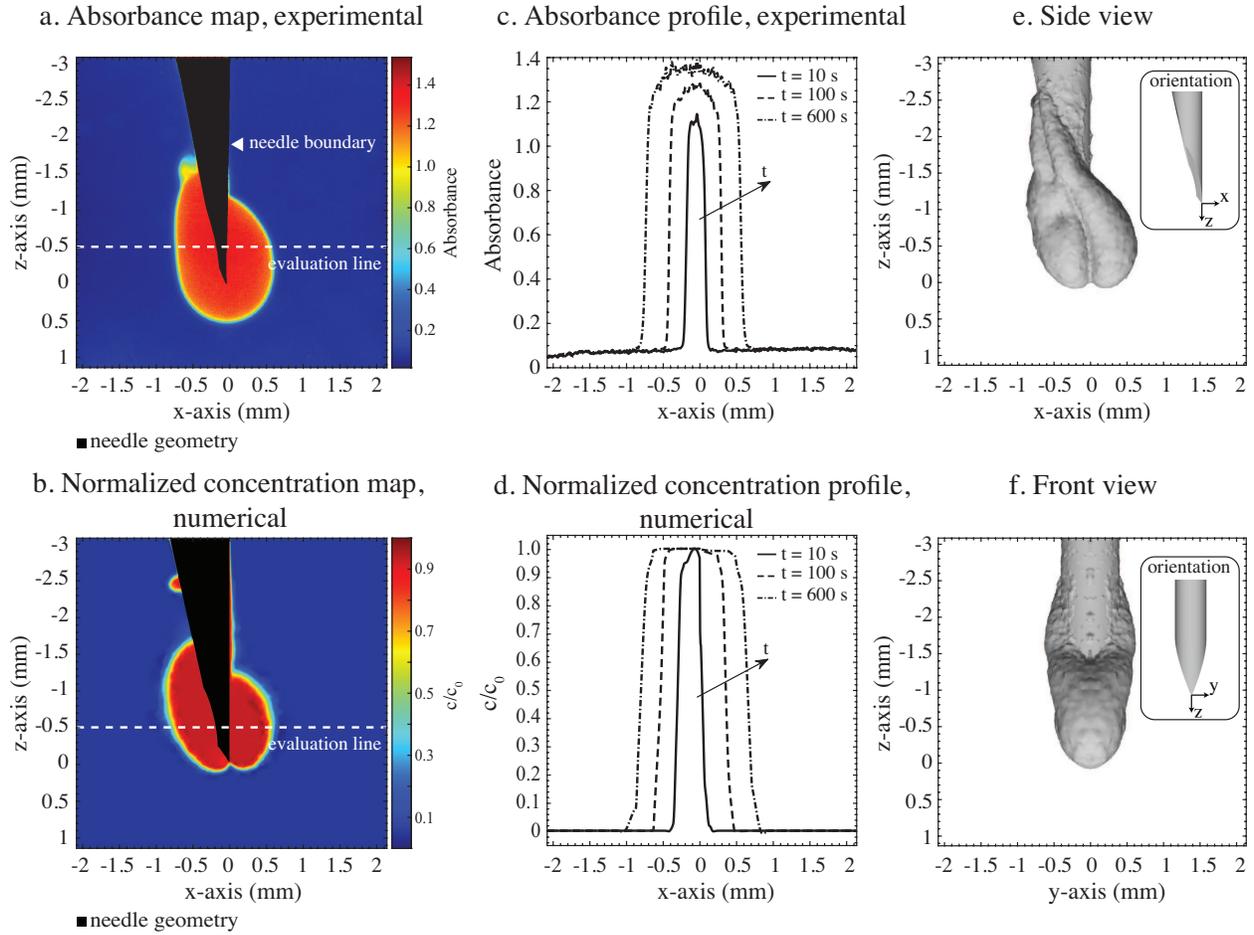}
\caption{\label{figure3}{(\textbf{a}) Optical absorbance (Beer-Lambert) image of a 0.5\% agarose gel sample exposed to \SI{10}{min} of ultrasound mediated delivery of 40\% (v/v, ink/deionized water) diluted ink and (\textbf{b}) cross-section of the simulated normalized concentration map. The needle tip geometry has been overlaid as a black polygon on the original images. (\textbf{c}) Absorbance profile and (\textbf{d}) simulated normalized concentration profile extrapolated from the evaluation line shown in \textbf{a} and \textbf{b}, respectively. These results suggest that the predicted dynamics of delivery manifested in a similar fashion during the experiments. (\textbf{e},\textbf{f}) Side and front view of the 3D reconstruction of the volume of delivery obtained form the simulation results. The shape suggests that the delivery mostly takes place at the needle tip with pronounced contribution along the xz-plane.}}
\end{figure*}

Fig. \ref{figure3}a depicts the absorbance map of a 0.5\% agarose specimen after the sonication was applied. The color intensity represents the light absorbance calculated as $A = \text{log}_{10}(I_0/I) $, where $I_0$ is the intensity of the incident light, considered as the background pixel intensity, and $I$ is the intensity of the light after it passed through the sample. A black polygon showing the needle contour has been overlapped to the absorbance map, since the absorbance values in that region are not representative of the tracer concentration. After \SI{10}{min} of sonication, the penetration front of the tracer extended to \SI{0.6}{mm} on both sides of the needle tip along the x-axis and to \SI{0.6}{mm} from the tip along the positive z-direction. Considering the simulated concentration map assessed on a cross-section passing through the needle center line and parallel to the xz-plane, the tracer expanded to \SI{0.5}{mm} radially from tip along the x-axis and to \SI{0.2}{mm} along the positive z-direction (Fig. \ref{figure3}b). The spreading velocity of the penetration front, evaluated on a line parallel to the x-axis with an offset of \SI{-0.5}{mm}, was on average \SI{2.2}{\micro\m\per\second}, calculated across a time window of \SI{100}{s}, while after \SI{200}{s} was found to spread at a constant velocity of \SI{0.6}{\micro\m\per\second} (Fig. \ref{figure3}c). Numerically, these velocities were calculated to be \SI{3.3}{\micro\m\per\second} and \SI{0.8}{\micro\m\per\second} across the same time windows (Fig. \ref{figure3}d). Fig. \ref{figure3}e and Fig. \ref{figure3}f show a 3D reconstruction of the simulated volume diffused by nanoparticles after \SI{10}{min} sonication, revealing that the delivery is localized at the needle tip. The side and front view suggest that the delivery action is predominantly within the xz-plane (Fig. \ref{figure3}e) and marginal on the yz-plane. This has to do with the flexural mode induced in the needle, which is made to oscillate sideways within the xz-plane, with greatest displacements at the tip. By this means, the tip acts as a dipole-like sound source, generating an acoustic intensity vector field with highest magnitudes located at the needle tip and with the same direction of needle motion.

\subsection{Numerical Simulation of Delivery in a Liver Tissue Model}
\begin{figure*}[ht]
\includegraphics[width=\textwidth]{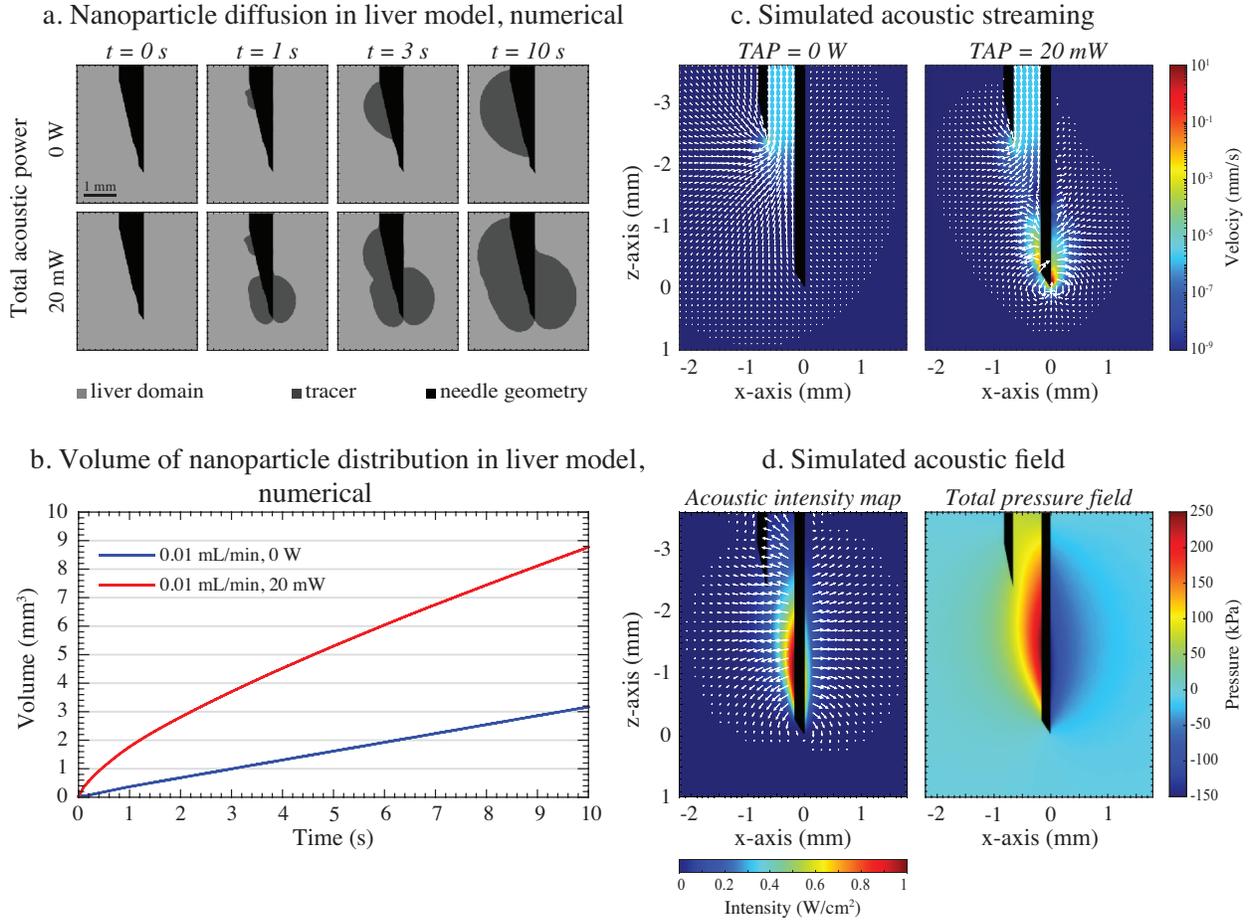}
\caption{\label{figure4}{Simulation study of ultrasound-mediated delivery in liver tissue, where permeability and porosity values were estimated using the Carman-Kozeny model. (\textbf{a}) The delivery effect was demonstrated by simulating the transport of the tracer around the needle tip under the action of the only applied flow rate of \SI{10}{\micro\liter\per\minute} (top row, TAP = \SI{0}{W}) as compared to when a TAP of \SI{20}{mW} was delivered by the needle. (\textbf{b}) The ultrasonic action of the needle exhibited a threefold increase of the volume diffused by the agent as compared to when no ultrasound was applied to the needle. (\textbf{b}) The enhanced delivery mechanism lies in the additional fluid flow pattern generated by ultrasound (acoustic streaming), which is extremely localized at the needle tip. This is also supported by (\textbf{d}) the simulated acoustic intensity map and pressure field, which reveal that the acoustic energy is highly concentrated in this region.}}
\end{figure*}

In order to demonstrate that the ultrasonic actuation of a hypodermic needle can bring contribution to a clinical scenario, we simulated the delivery action in a simplified model for liver \cite{Rezania2013}, which represents a target common in different clinical applications. The liver was modeled as a porous medium, whose permeability and porosity values were estimated using a Carman-Kozeny model \cite{Dullien1979}, which assumes flow through a bed of spherical objects with comparable radius to one of the hepatocytes ($\approx\SI{12}{\micro m}$). Simulations were carried out by considering a needle injection rate of \SI{10}{\micro\liter\per\minute} with delivered total acoustic power (TAP) of \SI{0}{W} and \SI{20}{mW} throughout \SI{10}{s}. The temporal variation of the volume diffused by the nanoparticles was investigated (Fig. \ref{figure4}a). When a normal injection was simulated, the drug was spread only on the side of the the needle opening. However, when a TAP of \SI{20}{mW} was employed, the drug spread more evenly around the needle tip (Fig. \ref{figure4}a). The volume of nanoparticle distribution increased at an average rate of \SI{0.3}{\cubic\mm\per\s}, when a flow rate of \SI{10}{\micro\liter\per\minute} was applied. Yet, in conjunction with US, it was estimated to be \SI{0.9}{\cubic\mm\per\s}. This resulted in a threefold increase in the total volume of delivery (Fig. \ref{figure4}b). 

The ultrasonic actuation of the needle contributed to a fluid velocity field highly localized at the very close proximity of the needle tip. The average stream velocity was \SI{1}{\mm\per\s} within \SI{100}{\micro m} from the tip along the x-axis, almost 5 orders of magnitude greater than the average velocity within the needle shaft caused by the externally applied flow rate (Fig. \ref{figure4}c). The simulated acoustic intensity map revealed that the maximum acoustic intensity is located near the tip, emanating outwards from the needle center line along the x-axis direction (Fig. \ref{figure4}d). Since the acoustic streaming is proportional to the acoustic intensity, higher stream velocities within the same plane were also generated. 

\section{Discussion}
These results suggest that  actuating a hypodermic needle with US allows one to enhance the delivery of nanoparticles and fluids in porous media, such as soft tissues. The resemblance between the numerical and experimental results of the transport velocities and tracer distribution within the sample indicates that convection and diffusion contribute to the transport mechanisms of nanoparticles in a porous medium. The tapered double-beveled structure of the needle provides greater needle displacements at its tip than elsewhere, therefore concentrating the acoustic intensity and amplifying the acoustic radiation pressure associated phenomena (\textit{i.e.} acoustic streaming) at this very location. Since most of the acoustic energy is confined at the proximity of the needle outlet (Fig. \ref{figure4}d), liquids or particles are allowed to be influenced by US, while being injected into the tissue, making this concept an interesting starting point for developing novel drug delivery systems combining needles and ultrasonic actuation.

Since the acoustic wavelength ($\sim\SI{45}{mm}$) at the employed frequency of \SI{33}{kHz} is considerably larger as than the diameter of the nanoparticles ($<$\SI{150}{nm}), the acoustic streaming is considered to be the main nonlinear acoustic phenomenon contributing to the delivery rather than acoustic radiation force directly pushing the particles. In fact, since the ratio between the acoustic streaming induced particle velocity and the ARF induced particle velocity is proportional to the square of the particle diameter, $d^2$, the ARF contribution can be neglected for particles smaller than \SI{150}{nm} in diameter \cite{Barnkob2012}. Potential cavitation activity or crack formation in the agarose specimen that might have contributed to enhancing the delivery effect were ruled out by reducing the TAP ($\sim$ \SI{20}{mW}). However, the sonication time had to be increased to \SI{10}{min} in order to observe a measurable effect.

Considering safety, the adopted frequency was relatively low compared to the MHz frequency typically employed in HIFU applications. Importantly, this limits the ultrasound-induced temperature increase. In fact, the acoustic absorption coefficient was estimated to be less than \SI{0.5}{\deci\bel\per\meter} at \SI{33}{kHz} in agarose. In addition, the US exposure of the agarose specimen did not exhibit any signs of cavitation or cavitation-induced liquefaction, as confirmed optically from the high-speed video footage. This suggests that the combination of the employed TAP and sonication duration is below the threshold for deleterious cavitation events. However, these effects are minimized further away from the needle, as the acoustic intensity geometrically decays by $1/r^2$, if assuming spherical wave propagation of the US wave.
Increased levels of TAP increase the probability of cavitation occurrences. If these phenomena were to take place in tissue, thermal and mechanical damage would occur in the regions where the acoustic intensity is greatest. Although this may represent a limitation when treating healthy tissues, conversely it might be beneficial for clinical procedures for tissue ablation, including solid organ cancers of the prostate \cite{Chen2016}, thyroid \cite{Pacella2013}, pancreas \cite{Larghi2021} and many more \cite{Piccioni2019}.
 
The idea of providing the acoustic energy locally with a conventional hypodermic needle to drive fluids and entities inside the body is first demonstrated in this study and with further improvements could be beneficial for drug delivery applications. This could be an alternative, minimally invasive and low-cost approach to expensive ultrasonic clinical methods, such as magnetic imaging-guided HIFU, used for image-guided delivery of drugs into a target location. Specifically, this method could add value \textit{e.g.} to the common liver cancer therapy of PEI \cite{ablationliver}. This technique consists of injecting ethanol into the tumor in order to achieve complete ablation of hepatocellular carcinoma. The PEI procedure is usually preferred to the non-invasive HIFU approach, when the tumor is located in the upper part of the liver. Since the rib cage and the lower part of the right lung partially cover this region anatomically, the formation of a proper ultrasonic focus in this area is difficult to achieve due to the poor acoustic window. The ultrasonic actuation of the needle would help in distributing ethanol locally and evenly within the tumor, by exerting a pushing force on the liquid surrounding the needle tip. This is supported by the numerical results, which showed a threefold increase in the volume of tissue diffused by the tracer when the needle was actuated for \SI{10}{s} at a TAP of \SI{20}{mW} under an applied flow rate of \SI{10}{\micro\liter\per\minute}, as compared to when no ultrasound was employed. This could potentially increase uptake of fluid caused by the enhanced delivery localized near the needle tip. These benefits could potentially bring added value of this technology for PEI and similar procedures, such as precision delivery of nanoparticles as personalised anticancer drugs \cite{Chen2021} and radiological contrast for discrete organ imaging in preparation for surgical resection \cite{Sier2021}. 

The presented technology, if optimized to its full potential, could be valuable for other medical applications. At high TAP levels, the needle vibrations are expected to induce cavitation and thermal effects. These can be of use in applications such as tumor ablation \cite{Zhou2011}, histotripsy or lithotripsy \cite{Yoshizawa2009}, where the objective is to cause mechanically pathological tissue destruction. The ability to deliver acoustic energy in a localized manner, while bypassing anatomical structures can be employed for the ablation of lung lesions, which are seldom treated with ultrasound due to presence of air that represents an obstacle for the US wave propagation. Moreover, in our recent study, we have demonstrated that the very same technology can be used to enhance the sample yield during fine needle aspiration biopsy (FNAB) \cite{Perra2021}, making this device clinically versatile and very promising for novel medical use. The results of the present study extent to broaden the understanding of fluid dynamics near the ultrasonically actuated medical needle, \textit{e.g.} in the context of ultrasonically enhanced biopsy. Such finding could open up an avenue of research application where FNAB and delivery of substances are combined, \textit{e.g.} delivery of contrast agent to enhance the needle visibility in ultrasound-guided FNAB.

\section{Conclusion}
We have demonstrated numerically that the ultrasonic actuation of a medical needle enhances delivery of nanoparticles and liquids in porous media, \textit{e.g.} soft tissues with porous structure, at the proximity of the needle tip. The delivery action was also observed experimentally in different agarose concentrations, with similar results to those obtained in the computational model. Based on the assumptions made in the numerical model and the similarity between the numerical and experimental results, it appears that enhanced convection by acoustic streaming is one of the main factors to the nanoparticle transport mechanisms in porous media. A simulation study considering the enhanced delivery in a liver model was also conducted in order to demonstrate its potential for biomedical research and clinical applications. As percutaneous injections are frequently conducted under US guidance in the clinic, we must eventually evaluate our technology in combination with simultaneous US imaging to optimise compatibility with established clinical procedures. To conclude, this method, if fully developed, could provide minimally invasive and localized drug delivery for small lesions in a safe, portable and cost-effective manner, while minimizing unnecessary drug exposure to adjacent organs.

\begin{acknowledgments}
We thank all members of the Medical Ultrasonics Laboratory (MEDUSA) at Aalto University for constructive discussions related to the topic. The Academy of Finland is acknowledged for financial support (grants 314286, 311586 and 335799).
\end{acknowledgments}

\section*{Author contributions statement}
All authors contributed to the design of the study, writing or reviewing the manuscript and have approved the final version of the manuscript. Emanuele Perra produced all data and conducted the data analysis.

\section*{Conflict of interest}
 Heikki J. Nieminen and Kenneth P.H. Pritzker have stock ownership in Swan Cytologics Inc., Toronto, ON, Canada and are inventors within the patent application WO2018000102A1. Emanuele Perra and Nick Hayward do not have any competing interests in relation to this work.  

\section*{Data availability}
The datasets are available upon request.

\section*{Code availability}
The codes are available upon request.

\bibliography{references}

%TABLES
\newpage
\begin{table*}[ht]
\caption{\label{Table1}A table of the general parameters of the numerical model at the ambient temperature of \SI{20}{\degreeCelsius}. The speed of sound in the agarose domain was assumed to be the same as in water.}
\resizebox{\textwidth}{!}{%
\begin{tabular}{lllllll}
\hline
Properties                        & unit              & water    & agarose (0.5\%) & agarose (1\%) & agarose (2\%) & liver    \\ \hline
Speed of sound, \textit{$c_0$}                 & \SI{}{\meter\per\second}    & 1482 \cite{Bilaniuk1993}     & 1482 \cite{Bilaniuk1993}            & 1482 \cite{Bilaniuk1993}          & 1482 \cite{Bilaniuk1993}          & 1575 \cite{Amin1989}     \\
Attenuation coefficient, $\alpha$ & \SI{}{\deci\bel\per\meter}   & \num{2.178e-4} \cite{Verma2005} & 0.165 \cite{Drakos2020}           & 0.330 \cite{Drakos2020}       & 0.495 \cite{Drakos2020}       & 1.39 \cite{Amin1989}     \\
Density, $\rho$                   & \SI{}{\kg\per\cubic\m} & 998.2 \cite{Pahk2019}     & 1003 \cite{Laurent1967}  & 1006 \cite{Laurent1967}         & 1012 \cite{Laurent1967}         & 1060 \cite{Pahk2019}     \\
Porosity, $\epsilon$              & -                 & -        & 0.995 \cite{Johnson1996}            & 0.990    \cite{Johnson1996}      & 0.980 \cite{Johnson1996}        & 0.476 \cite{Rezania2013} \\
Permeability, $\kappa$            & \SI{}{m^2}        & -        & \num{2.27e-15} \cite{Debbaut2014}      & \num{1.31e-15} \cite{Debbaut2014}     & \num{1.9e-16} \cite{Debbaut2014}      & \num{1.26e-12} \cite{Dullien1979} \\ \hline
\end{tabular}%
}
\end{table*}

%FIGURES
%\newpage

%\newpage

%\newpage

%\newpage
\end{document}